\documentclass[twocolumn]{article}
\usepackage[pdftex]{graphicx}
\usepackage{authblk}

\usepackage{bm}
\usepackage{siunitx}[=v2]
\usepackage{physics}
\usepackage{amsmath}
\usepackage[version=4]{mhchem}
\usepackage[margin=25truemm]{geometry}

\title{Spontaneous Anomalous Hall Effect at Room Temperature in Antiferromagnetic Material NbMnAs}

\author[1]{Yuki Arai}
\author[2]{Junichi Hayashi}
\author[2]{Keiki Takeda}
\author[1]{Hideki Tou}
\author[1]{Eiichi Matsuoka}
\author[1]{Hitoshi Sugawara}
\author[1]{Hisashi Kotegawa}
\affil[1]{Department of Physics, Kobe University, Kobe, Hyogo 657-8501, Japan}
\affil[2]{Muroran Institute of Technology, Muroran, Hokkaido 050-8585, Japan}

\begin{document}
\maketitle
\twocolumn[
\begin{abstract}
Recent studies have shown that certain antiferromagnetic (AFM) materials with the same symmetry breaking as ferromagnets can generate sufficiently large ferromagnetic (FM) responses.
Here, we report that the new AFM material NbMnAs exhibits a large anomalous Hall effect (AHE) at zero field and at room temperature, despite having only a small net magnetization.
A polycrystalline sample of NbMnAs, likely close to stoichiometric composition, exhibited an AFM state with a small spontaneous magnetization of approximately $6 \times 10^{-3} \mu_{\rm B}$/Mn and the AHE below $T_{\rm N}=\SI{354}{K}$.
In contrast, single crystals of NbMnAs obtained by a flux method exhibited a deficiency at the As site, {which resulted} in a decrease in $T_{\rm N}$ and an increase in spontaneous magnetization.
Although improvement of the single-crystal growth is still required, our study reveals that NbMnAs is a novel material capable of exhibiting significant FM responses derived from antiferromagnetism at room temperature.
\newline
\end{abstract}
]

The anomalous Hall effect (AHE) refers to a spontaneous Hall effect which occurs even in the absence of an external magnetic field.
It arises from broken symmetries in a material leading to a nonzero off-diagonal component in the electrical conductivity tensor.
In recent years, it has been revealed that the origin of these symmetry breaking is not limited to magnetic fields or magnetization; antiferromagnetic (AFM) structures, symmetrically equivalent to ferromagnetic (FM) states, can also give rise to the same effect \cite{Chen14,Suzuki17}.
Indeed, the AHE has been experimentally observed in several materials possessing AFM ordering \cite{Nakatsuji2015,Kiyohara16,Nayak16_MnGe,Ghimire18_CoNbS,Akiba20_alpha-Mn,Park22_CoTaS,Kotegawa_NbMnP,Gonzalez_MnTe,Kotegawa_TaMnP,Takagi_FeS,Kotegawa_CeCuGe_CePdGe}.
In most cases, the AHE is driven by a dissipationless intrinsic mechanism described by Berry curvature in momentum space, which implies that the effect persists irrespective of sample quality \cite{nagaosa_anomalous_2010}.
These materials also exhibit other FM responses, such as the anomalous Nernst effect, the magneto-optical Kerr effect, and the magnetic spin Hall effect, which make them promising candidates for future spintronics and thermoelectric technologies. \cite{Nakatsuji2022topological}.
{While research on thin films and devices plays an important role in the development of applications, including enhanced output, diversified responses, and controllability of magnetic domains \cite{Matsuda_2020,Takeuchi,Hu,Jeon,Shukla}, studies of bulk materials are essential for understanding fundamental physical properties, and the exploration of new materials is critically important for pursuing higher transition temperatures and larger responses.}
The AHE has been observed at room temperature in only a few materials as bulk\cite{Nakatsuji2015,Kiyohara16,Nayak16_MnGe,Gonzalez_MnTe,Takagi_FeS}, and large anomalous Hall conductivities (AHCs) comparable to those of ferromagnets has been reported exclusively in Mn$_3$Sn and Mn$_3$Ge in bulk crystal forms \cite{Nakatsuji2015,Kiyohara16}.
{Exploring new materials that exhibit large AHCs at room temperature is of considerable importance but remains extremely challenging.}
\begin{figure}[h]
\includegraphics[scale=0.62]{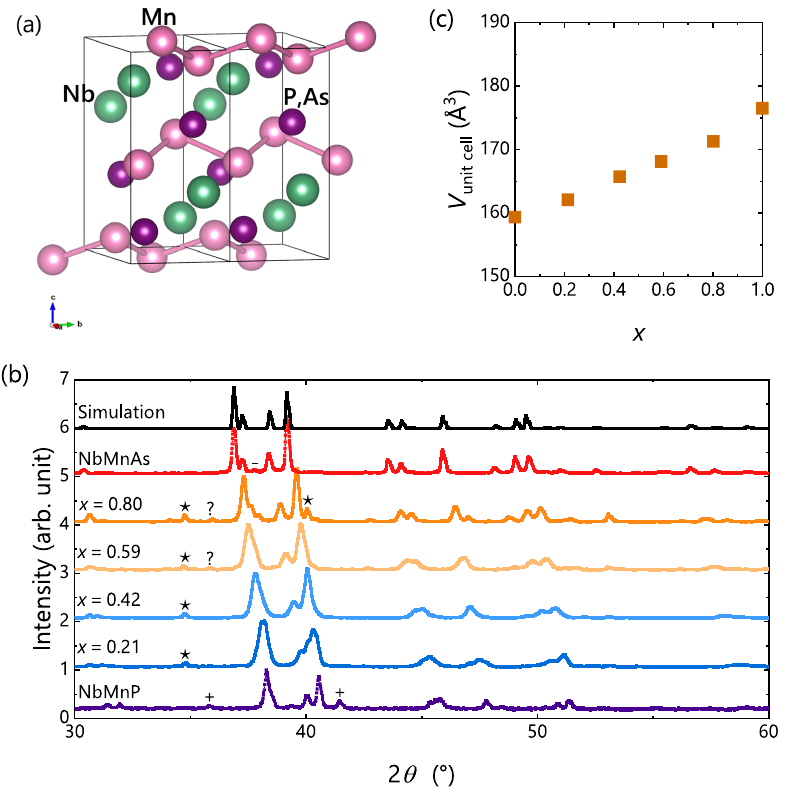}
\caption{(a) Orthorhombic crystal structure of NbMnP and NbMnAs. The Mn atoms form zigzag chains along the $b$-axis. (b) XRD patterns for NbMnP$_{1-x}$As$_{x}$. Each $x$ was determined by XRF measurements. The black line represents the simulated pattern for NbMnAs, which has the same space group as NbMnP. The symbols indicate impurity phases: $\star$: NbAs, +: NbP, -: Nb, and ?: {an unknown phase}. (c) Substitution dependence of the unit-cell volume. The continuous increase in volume with increasing As content was observed.}
\label{fig_2}
\end{figure}

Recently, our group found that the AFM metal NbMnP exhibits a large AHC below $T_\mathrm{N}=\SI{244}{K}$ \cite{Kotegawa_NbMnP,Arai_2024}.
NbMnP crystallizes in a TiNiSi-type orthorhombic structure with the space group $Pnma$, as shown in Fig. \ref{fig_2}(a){, illustrated using VESTA\cite{Lomnitskaya,VESTA}.}
The magnetic structure, characterized by a propagation vector $q=0$, belongs to the magnetic point group $m'm2'$ \cite{Matsuda_2022,Kotegawa_NbMnP}, which results from a linear combination of $mm'm'$ and $mm'm$.
The $mm'm'$ symmetry allows both AFM and FM states; therefore, the AFM state with this symmetry can generate the AHE.
This symmetry also permits a spontaneous magnetization, and it is $2\times 10^{-3} \mu_\mathrm{B}$/Mn for NbMnP against the AFM ordered moment of 1.2 $\mu_\mathrm{B}$/Mn \cite{Matsuda_2022}.
Among TiNiSi-type orthorhombic structure containing Mn, TaMnP also exhibits a large AHC derived from the AFM state with an $mm'm'$ component below $T_{\rm N}=\SIrange[range-phrase={\textendash},range-units=single]{220}{240}{K}$ \cite{Kotegawa_NbMnP}.
In contrast, ZrMnP and HfMnP show typical FM ordering \cite{Lamichhane_ZrMnP}.
\ce{YMnGe} exhibits a $q=0$ AFM state, but its symmetry does not allow the AHE \cite{Klosek_YMnGe}, while LuMnSi shows a spiral AFM state \cite{Venturini_LuMnSi}.
These results indicate that the change in the valence electrons significantly affects the type of magnetic ordering.
In this context, substituting P with As is expected to preserve the AFM state that generates the AHE, while potentially modifying the transition temperature via an increase in the unit-cell volume.
However, it should be noted that the existence of a fully substituted compound, NbMnAs, had not been reported yet.

In this study, we found that substituting P with As increases the unit-cell volume without altering the crystal structure of the $Pnma$ space group, and that NbMnAs can exist stably.
The $T_{\rm N}$ increased monotonically with increasing As content, reaching $\SI{354}{K}$ for NbMnAs.
The small spontaneous magnetization of $6\times 10^{-3} \mu_\mathrm{B}$/Mn for NbMnAs (comparable to $2\times 10^{-3} \mu_\mathrm{B}$/Mn for NbMnP) indicates that the magnetically ordered state is AFM.
Below $T_{\rm N}$, NbMnAs exhibits a large AHE even at zero magnetic field, which means that it is a new AFM material with a large AHC at room temperature.

%
%
%
Polycrystalline samples of NbMnP$_{1-x}$As$_{x}$ (nominal composition: $x=0.2, 0.4, 0.6, 0.8, 1$) were prepared by a solid-state reaction.
For $x=0.2\textendash 0.8$, mixtures of Nb powder, Mn powder, P flakes, and As flakes with a molar ratio of $1:1:1-x:x$ were weighed and sealed in evacuated quartz ampoules.
The ampoules were gradually heated in several steps from \SI{600}{\degreeCelsius} to \SI{800}{\degreeCelsius} to promote the reaction of As, and then further heated at \SI{950}{\degreeCelsius} for 100 hours.
We repeated pulverization and annealing several times to achieve homogenization of the sample.
Polycrystalline samples of NbMnAs were synthesized in a similar manner to NbMnP$_{1-x}$As$_{x}$, except that Nb powder, Mn powder, and As flakes were used with a molar ratio of $1.1:1.1:1$.
The excess Nb and Mn were effective in reducing impurity phases, although a small amount of unreacted Nb remained in the final sample.
Single crystals of NbMnAs were grown by the Ga--In flux method using Nb powder, Mn chips, As flakes, Ga chips, and In chips with a molar ratio of $1:3:1:8:12$.
The materials were placed in an alumina crucible and sealed in an evacuated quartz ampoule.
The ampoule was gradually heated to \SI{1200}{\degreeCelsius}, held at that temperature for 6 hours, and then slowly cooled to \SI{750}{\degreeCelsius}.
Small needle-like single crystals were obtained after centrifugation.
Powder X-ray diffraction (XRD) measurements were conducted using a MiniFlex diffractometer (Rigaku) to confirm the crystal structure of the synthesized samples.
For the single crystal, the crystal symmetry, lattice parameters, and site occupancies were determined by single-crystal XRD measurements using a Rigaku Saturn724 diffractometer.
The element composition of the samples were evaluated by X-ray fluorescence (XRF) spectroscopy using a JSX-1000S (JEOL).
The crystallographic orientation of the single crystal was determined by the Laue diffraction method.
For electrical transport measurements, four gold wires were bonded to each samples: a polycrystalline pellet or a single crystal.
We antisymmetrized the Hall resistivity with respect to the magnetic field to remove the longitudinal component arising from potential contact misalignments.
We measured magnetization using a commercial SQUID magnetometer (MPMS, Quantum Design) on a polycrystalline pellet or multiple single crystals arranged perpendicular to the $b$ axis.
%

\begin{table}[h]
\caption{{Crystal structure and structural parameters of single-crystal NbMnAs grown by the Ga--In flux method, determined by single-crystal XRD measurement. The estimated occupancies at each site are also shown.}}
\label{t1}
\begin{center}
\begin{tabular}{|l|c|}
\hline
Temperature & 293 K \\
Formula & NbMnAs$_{0.85}$ \\
Crystal system & orthorhombic \\
Space group & $Pnma$ (no.62) \\
$a$ (\AA) & 6.3726(4) \\
$b$ (\AA) & 3.6406(3) \\
$c$ (\AA) & 7.5427(7) \\
$V$ (\AA$^3$) & 174.99(2) \\
$Z$ & 4 \\
Independent reflections & 280 \\
Residual factor $R1$ & 0.0277 \\
$wR2$ & 0.0660 \\
\hline
\end{tabular}
\vspace{.5cm}
\begin{tabular}{cccc}
\\ \hline
Atom & Nb & Mn & As \\
\hline\hline
$x$ & 0.03606 & 0.13896 & 0.26173 \\
$y$ & 0.250000 & 0.250000 & 0.250000 \\
$z$ & 0.67134 & 0.05956 & 0.36742 \\
Occup. & 1 & 1 & 0.852 \\
$U (\si{\angstrom^2})$ & 0.006& 0.005 & 0.006 \\
\hline
\end{tabular}
\end{center}
\end{table}

\begin{table*}[h]
\caption{Comparison of the crystal structure, $T_{\rm N}$, element compositions among NbMnAs (polycrystal), NbMnAs (single crystal), and NbMnP. {The chemical composition of each sample was estimated using XRF and single-crystal XRD measurements.} }
\label{t2}
\begin{center}
\begin{tabular}{|l||c|c|c|}
\hline
compound & NbMnAs & NbMnAs & NbMnP \\
form & polycrystal & single crystal & single crystal \\
space group & $Pnma$ & $Pnma$ & $Pnma$ \\
$T_{\rm N}$ & 354 & \numrange[range-phrase={--}]{290}{340} & 244 \\
composition (XRF) & Nb$_{0.92}$Mn$_{1.07}$As$_{0.96}$ & Nb$_{0.98}$Mn$_{1.01}$As$_{0.72}$ & Nb$_{0.91}$Mn$_{1.02}$P$_{1.07}$ \\
composition (single-crystal XRD) & -- & NbMnAs$_{0.85}$ & Nb$_{0.95}$MnP \\
$a$ (\AA) & 6.3913 & 6.3726 & 6.1899 \\
$b$ (\AA) & 3.6793 & 3.6406 & 3.5478 \\
$c$ (\AA) & 7.5468 & 7.5427 & 7.2380 \\
$V$ (\AA$^3$) & 177.5 & 174.99 & 158.95 \\
\hline
\end{tabular}
\end{center}
\end{table*}

%
%

Fig. \ref{fig_2}(b) shows the powder XRD patterns for the polycrystalline samples of NbMnP$_{1-x}$As$_{x}$.
The substitution ratios $x$ were estimated by the XRF measurements, and the differences between the nominal and actual compositions were within 5\%.
Each diffraction peak shifts toward lower angles with increasing As content, indicating lattice expansion.
As shown in the figure, the diffraction pattern of NbMnAs is also well explained by the TiNiSi-type structure, which is the same as that of NbMnP.
Each lattice constant was determined from these diffraction patterns, and the unit-cell volumes are plotted in Fig. \ref{fig_2}(c).
The monotonic increase in volume supports that the crystal symmetry remains unchanged from the original $Pnma$ structure.
For the substituted samples, several peaks corresponding to impurity phases such as NbP or NbAs were detected, whereas the polycrystalline NbMnAs was more homogeneous.
A small amount of residual Nb was detected due to the excess Nb used in the starting materials.
The crystal structure of the single crystal of NbMnAs was examined by single-crystal XRD measurements, and the results are summarized in Table \ref{t1}.
The space group was confirmed to be $Pnma$, as expected.
{An obvious deficiency at the As site ($4c$ site) was observed, as the occupancy was estimated to be 85.2\%.
This deficiency was also confirmed by XRF measurements, as shown in Table \ref{t2}, where the comparison among the polycrystalline sample and NbMnP is also listed.
In the XRF measurements, the polycrystalline sample is close to stoichiometric composition, whereas the occupancy of the As site in the present single crystal is approximately $72\%$. While the occupancy shows some variation depending on the measurement method, a significant deficiency at the As site is evident.}
The difference in volume between the polycrystalline and the single-crystal samples is likely due to the As deficiency.

\begin{figure}[h]
\begin{center}
\includegraphics[scale=0.62]{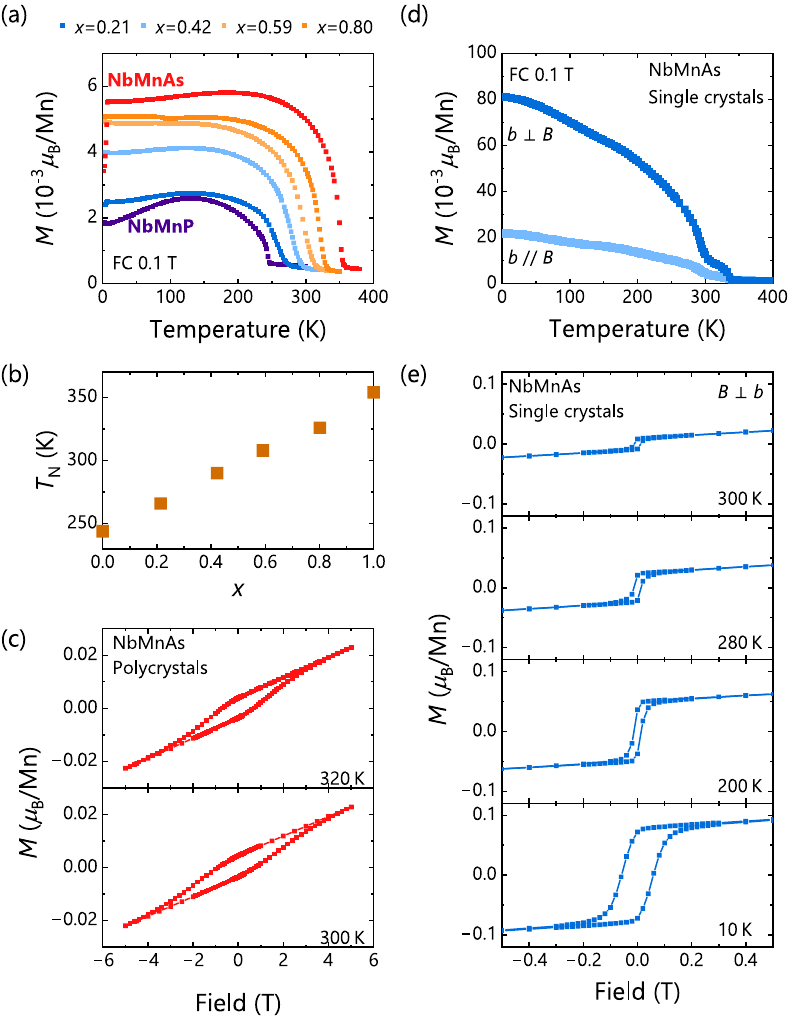}
\end{center}
\caption{(Color online) (a) Temperature dependence of the magnetization at \SI{0.1}{T}, (b) substitution dependence of $T_{\rm N}$ and (c) hysteresis loops for the polycrystalline samples. A small but distinct increase in magnetization appears below $T_{\rm N}$. For NbMnAs, a drop in magnetization is observed due to the superconductivity of the Nb impurity. The continuous increase in $T_{\rm N}$, as well as the unit-cell volume, suggests that the magnetically ordered states of NbMnP and NbMnAs are described by the same symmetry. (d) Temperature dependence of the magnetization at \SI{0.1}{T} and (e) hysteresis loops for the single crystals. A distribution of $T_{\rm N}$ was observed, ranging from \SIrange{290}{340}{K}. The spontaneous magnetization is about ten times larger than that of the polycrystalline sample, which is presumed to originate from As deficiency.
}
\label{fig_3}
\end{figure}


To determine $T_\mathrm{N}$ of the obtained samples, magnetization measurements were performed at \SI{0.1}{T} as a function of temperature.
The results for the polycrystalline samples of NbMnP$_{1-x}$As$_{x}$ are shown in Fig. \ref{fig_3}(a).
All compounds show an increase in magnetization below their transition temperatures.
This behavior originates from the appearance of a weak FM component, which is expected to emerge along the $a$-axis via the Dzyaloshinskii--Moriya (DM) interaction, because the symmetry of the AFM state in NbMnP is equivalent to that of an FM state along the $a$-axis \cite{Kotegawa_NbMnP,Arai_2024}.
The dependence of $T_{\rm N}$ on the substitution ratio $x$ is plotted in Fig. \ref{fig_3}(b).
$T_{\rm N}$ increases monotonically with increasing $x$, reaching \SI{354}{K} for NbMnAs.
This monotonic change and the presence of a weak FM component indicate that the AFM state of NbMnAs is symmetrically equivalent to that of NbMnP.
Fig. \ref{fig_3}(c) shows the magnetic-field dependence of the magnetization below $T_\mathrm{N}$ for polycrystalline NbMnAs.
An external magnetic field must be applied along the $a$-axis to reverse the magnetic domains.
However, because the sample is polycrystalline, a large magnetic field is required to reverse the domains throughout the entire sample; therefore, it was difficult to achieve complete domain reversal below \SI{300}{K} under \SI{8}{T}.
At \SI{300}{K}, the spontaneous magnetization of NbMnAs was $4\times10^{-3}\mu_\mathrm{B}$/Mn, which is almost consistent with the magnetization measured at \SI{0.1}{T}.
Thus, Fig. \ref{fig_3}(a) suggests that the spontaneous magnetization of NbMnAs at low temperatures is $6\times 10^{-3}\mu_\mathrm{B}$/Mn, which is approximately three times larger than that of NbMnP.

\begin{figure*}[h]
\begin{center}
\includegraphics[scale=0.62]{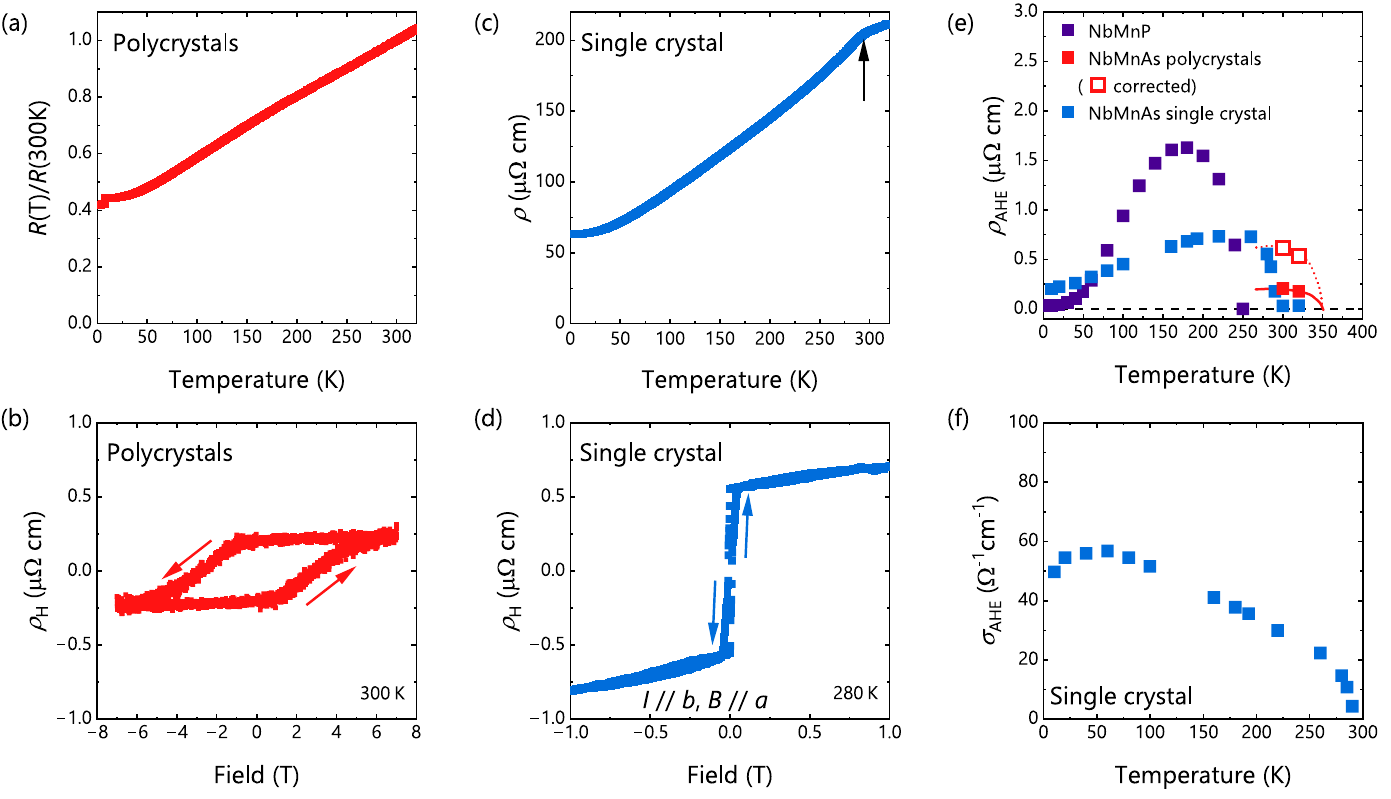}
\end{center}
\caption{(a) Temperature dependence of a proportion of the electrical resistance $R(T)$ in $R(\SI{300}{K})$ for the polycrystalline NbMnAs with $T_{\rm N}=354$ K. (b) Hysteresis loop of $\rho_\mathrm{H}$ for the polycrystalline NbMnAs. (c) Temperature dependence of $\rho$ for the polycrystalline and single-crystal NbMnAs. Red lines are guide to data points of polycrystals. (d) Hysteresis loop of $\rho_\mathrm{H}$ for the single-crystal NbMnAs. (e) Temperature dependence of $\rho_\mathrm{AHE}$ for the polycrystalline and single-crystal NbMnAs and single-crystal NbMnP.(f) Temperature dependence of $\sigma_\mathrm{AHE}$ for the single-crystal NbMnAs.
}
\label{fig_4}
\end{figure*}

The temperature dependence of the magnetization at \SI{0.1}{T} for multiple single crystals arranged {parallel and} perpendicular to the $b$ axis is shown in Fig. \ref{fig_3}(d).
{The magnetization of single crystals clearly shows the anisotropy.}
A slight increase in magnetization was observed below approximately \SI{340}{K}, while a distinct kink appeared at \SIrange[range-phrase={\textendash},range-units=single]{290}{300}{K}.
{Although the possibility of a successive phase transition cannot be completely ruled out, such behavior was not observed in the polycrystalline samples. 
In addition, multiple samples were used in the magnetization measurements, and we found that $T_{\rm N}$ of the single-crystal samples exhibited considerable variation even within a single batch. 
Therefore, we infer that the two-step anomaly observed in the magnetization originates from a distribution of $T_{\rm N}$ among different single crystals.}
Another difference between the polycrystalline and single-crystal samples is the magnitude of the FM component.
As shown in Fig. \ref{fig_3}(e), the spontaneous magnetization at low temperatures was $8\times 10^{-2}\mu_\mathrm{B}$/Mn, which is more than an order of magnitude larger than that of the polycrystalline sample.
{If the magnetic structures of NbMnAs and NbMnP are described by the same magnetic point group, the magnetization along the $b$ axis cannot be explained.
Since no indication of a change in the magnetic point group is observed in the polycrystalline samples NbMnP$_{1-x}$As$_x$, the magnetization along the $b$ axis is inferred to originate primarily from poor quality of the single crystals.
Similar behavior has been observed in slightly off-stoichiometric TaMnP \cite{Kotegawa_TaMnP}.
In addition, the overall increase in the magnitude of the magnetization is also likely attributable to the sample quality.
}

The presumption that the magnetic structures of NbMnP and NbMnAs are symmetrically equivalent is confirmed by the observation of AHE.
Figure \ref{fig_4}(a) shows the temperature dependence of the electrical resistance for the polycrystalline sample with $T_{\rm N}=\SI{354}{K} $, measured up to \SI{320}{K}.
{Since the electrical resistivity of the polycrystalline sample was estimated to be significantly larger than that of a single crystal and the influence of grain boundaries cannot be excluded, the resistivity of the polycrystalline sample is normalized at room temperature. Note that it was $\rho(\SI{300}{K}) = \SI{1.0e3}{\micro\ohm \centi\meter}$.
In the resistivity measurement,} we confirmed that $T_{\rm N}$ exceeds \SI{320}{K}.
The small drop in $R(T)$ at low temperatures originates from the superconductivity of the Nb impurity corresponding to the magnetization data.
The residual resistivity ratio (RRR) was slightly greater than 2.
Figure \ref{fig_4}(b) shows the field dependence of the Hall resistivity, $\rho_\mathrm{H}$, for the polycrystalline sample at \SI{300}{K}.
{In general, the Hall effect does not include contributions from scattering; therefore, the Hall resistivity is presented for the polycrystalline sample.}
The $\rho_\mathrm{H}$ exhibited a clear hysteresis similar to that of the magnetization and retains a nonzero value at zero field.
Figure \ref{fig_4}(c) shows that the temperature dependence of $\rho$ for a single crystal.
This crystal exhibits $T_{\rm N}$ at \SI{294}{K}, which is {likely} reduced from \SI{354}{K} owing to As deficiency.
At \SI{280}{K} below $T_{\rm N}$, the single crystal also shows a clear hysteresis in $\rho_{\rm H} =\rho_{yz}$, as presented in Fig. \ref{fig_4}(d).
In this measurement, the current was applied along the $b$($y$)-axis, and the voltage was measured along the $c$($z$)-axis, and the magnetic field was applied along the $a$($x$)-axis.
The temperature dependences of $\rho_{\rm AHE}$, which is $\rho_{\rm H}$ at zero field, are summarized in Fig. \ref{fig_4}(e).
The magnitude of $\rho_{\rm AHE}$ in the polycrystalline sample is somewhat smaller than that in the single crystal, when compared at similar $T/T_{\rm N}$ values.
This difference is attributed to the averaging of the Hall voltage in the polycrystalline sample.
Assuming that the magnetic structure of NbMnAs contains the magnetic point group $mm'm'$ component, the Hall resistivity tensor is given by $\rho_{yx}=\rho_{xz}=0$ and $\rho_{zy}\neq0$.
For a polycrystalline sample, the observed Hall resistivity is expected to be $\rho_{\rm H}^{\rm obs}=(\rho_{yx}+\rho_{zy}+\rho_{xy})/3=\rho_{zy}/3$.
The corrected values, $\rho_{zy}=3\rho_{\rm AHE}^{\rm obs}$, are plotted as open symbols in Fig. \ref{fig_4}(e).
Now, they show similar value to those of the single crystal at comparable $T/T_{\rm N}$.
Figure \ref{fig_4}(f) shows the temperature dependence of the AHC, $\sigma_\mathrm{AHE}$, for the single crystal, evaluated as $\sigma_\mathrm{AHE} = \rho_\mathrm{AHE} / \rho^2$.
$\sigma_\mathrm{AHE}$ increases below $T_\mathrm{N}$ and reaches approximately \SI{60}{\ohm^{-1} cm^{-1}} at low temperatures.
This value is smaller than \SIrange[range-phrase={\textendash},range-units=single]{230}{450}{\per\ohm\per\centi\meter} observed for \ce{NbMnP} and \SI{380}{\per\ohm\per\centi\meter} for \ce{Mn3Ge} \cite{Kotegawa_NbMnP,Arai_2024,Kiyohara16,Nayak16_MnGe}, and about half of \SI{130}{\per\ohm\per\centi\meter} for \ce{Mn3Sn} \cite{Nakatsuji2015}.
Since the AHC of NbMnP has been explained by the theoretical calculations involving Berry curvature in momentum space \cite{Kotegawa_NbMnP}, the topological properties of NbMnAs can likely be clarified through band-structure calculations.
It is worth noting that MnTe and FeS also exhibit AHEs derived from AFM states at room temperature; however, their AHCs are estimated less than \SI{1}{\ohm^{-1}cm^{-1}} \cite{Gonzalez_MnTe,Takagi_FeS}, probably due to their semiconducting properties.
Therefore, NbMnAs can be regarded as a novel AFM material that generates a large AHC at room temperature, comparable to those of Mn$_3$Sn and Mn$_3$Ge.

To summarize, we have discovered a new AFM compound, NbMnAs, which crystallizes in an orthorhombic structure with the space group $Pnma$ and exhibits a large AHE below $T_\mathrm{N} = \SI{354}{K}$.
The observation of the AHE indicates that the magnetic structure of NbMnAs belongs to a magnetic point group that permits other FM responses.
Optical and thermal counterparts of the AHE --- namely, the magneto-optical Kerr effect and the anomalous Nernst effect --- are also expected to appear at room temperature. Although the present single crystal exhibits a non-negligible As deficiency, NbMnAs represents a promising candidate material for future spintronic and thermoelectric technologies.

\section*{Acknowledgements}
This study was supported by JST SPRING (Grant Number JPMJSP2148), the JSPS KAKENHI Grant Nos. 23H04871, 25K00947, and 25K21684, Iketani Science and Technology Foundation, Hyogo Science and Technology Association, and Murata Science Foundation.
We are grateful to Research Facility Center for Science and Technology of Kobe University for supporting the experiments involving MPMS and liquid helium and nitrogen.

\end{document}